\documentclass[a4j,12pt]{article}
\usepackage[dvips]{graphicx}
\begin{document}
\def\thefootnote{\arabic{footnote}}

\noindent
{\bf \large A mathematically rigorous approach raises questions concerning the generalized Hartman effect}

\vskip 5mm

\noindent
Shoju Kudaka $^a$ and Shuichi Matsumoto $^{b,}$\footnote{ Corresponding author. {\it E-mail address}: shuichi@edu.u-ryukyu.ac.jp} 

\noindent
{\it $^a$Department of Physics, University of the Ryukyus, Okinawa 903-0129, Japan}\\
{\it $^b$Department of Mathematics, University of the Ryukyus, Okinawa 903-0129, Japan}
\vfill

\noindent
{\bf Abstract}: With reference to a particle tunneling through two successive barriers, it seems to have been generally accepted that the tunneling time does not depend on the separation distance between the barriers. This phenomenon has been called the {\it generalized Hartman effect}. In this letter, we point out a lack of mathematical rigour in the reasoning by which this effect was deduced about ten years ago. A mathematically rigorous treatment shows us that the tunneling time does indeed depend on the length of the free space between the barriers.


\noindent
{\bf Pacs}: 03.65.Xp

\vfill

Consider the integral 
\begin{equation}
I=\int G(E)e^{i\theta (E)}dE, \label{eq:saisyonosiki}
\end{equation}
where $G(E)$ and $\theta (E)$ are real valued functions of a real variable $E$, and $G(E)$ is assumed to be {\it a sharply peaked function such as a Gaussian centered at a mean value}. Denote the deviation of $G(E)$ about its mean value $E_1$ as $\delta E$.

Let
\begin{equation} 
\theta (E)=\theta (E_1)+\theta '(E_1)(E-E_1)+O[(E-E_1)^2]
\end{equation}
be a Taylor expansion of $\theta (E)$. In the case that we can neglect the third term of the right side above for some reason, the integral $I$ can be rewritten as 
\begin{equation}
I=e^{i\theta (E_1)}\int G(E)e^{i\theta '(E_1)(E-E_1)}dE.
\end{equation}
If 
\begin{equation}
\theta '(E_1)\delta E\gg 1,
\end{equation}
then the term $e^{i\theta '(E_1)(E-E_1)}$ rapidly oscillates as $E$ varies around $E_1$, and $I$ approximates to zero. In contrast, if $\theta '(E_1)\delta E\approx 0$, especially if 
\begin{equation}
\theta '(E_1)=0,
\end{equation}
then the oscillation is suppressed and $I$ results in an appreciable value.

\begin{figure}
\begin{center}
\includegraphics[height=4cm,width=14cm,keepaspectratio,clip]{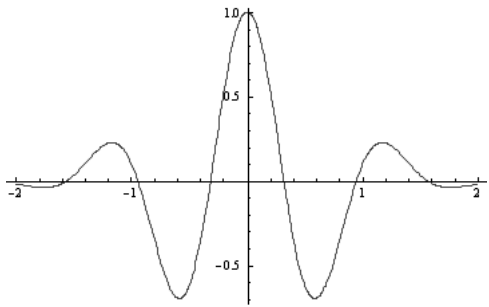}
\end{center}
\caption{ $G(E)$ given by Eq. (\ref{eq:myounagE}) \hskip 5mm ($\tau =5$)}
\label{figure:expcos}
\end{figure}

This is the essence of the stationary phase method (SPM). Here, we have to emphasize that the condition indicated in italics for the function $G(E)$ is very important in the reasoning. For instance, consider the function 
\begin{equation}
G(E)=e^{-E^2}\cos (\tau E) \label{eq:myounagE}
\end{equation}
where $\tau $ is a real parameter. Fig. \ref{figure:expcos} shows the graph of $G(E)$ for $\tau =5$. In that $G(E)$ firstly takes both positive and negative values and secondly has multiple extrema, this function does not satisfy the condition indicated in italics. 

Using $G(E)$ given by Eq. (\ref{eq:myounagE}), consider the integral 
\begin{equation}
I(t)=\int G(E)e^{itE}dE
\end{equation}
where $t$ is a real parameter. If we apply the SPM to this integral, since   
\begin{equation}
{d\over {dE}}tE=t,
\end{equation}
we are led to the conclusion that the value of $I(t)$ is appreciable only for $t=0$ and is negligible for sufficiently large $t$. 

\begin{figure}
\begin{center}
\includegraphics[height=4cm,width=14cm,keepaspectratio,clip]{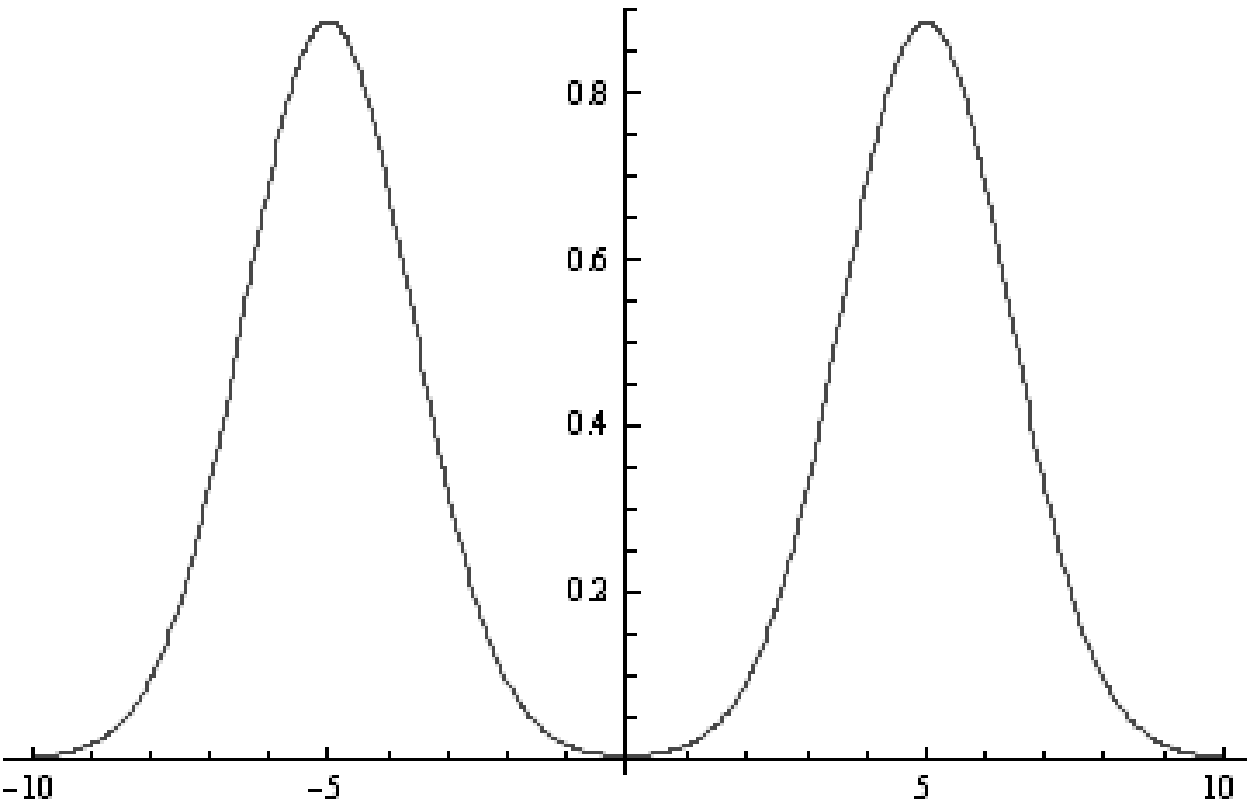}
\end{center}
\caption{$I(t)$ given by Eq. (\ref{eq:Itnokatachi}) \hskip 5mm ($\tau =5$)}
\label{figure:cosIt}
\end{figure}

However, since 
\begin{equation}
G(E)=e^{-E^2}\left( e^{i\tau E}+e^{-i\tau E}\right) /2,
\end{equation}
$I(t)$ can be expressed as  
\begin{equation}
I(t)={{\sqrt {\pi }}\over 2}\left( e^{-(t+\tau )^2/4}+e^{-(t-\tau )^2/4}\right) .\label{eq:Itnokatachi}
\end{equation}
The $t$ for which the value of $I(t)$ is appreciable is therefore not $t=0$ but $t=\pm \tau $. The graph of $I(t)$ as a function of $t$ is shown in Fig. \ref{figure:cosIt} for $\tau =5$. 

Thus, when we use the SPM for the integral (\ref{eq:saisyonosiki}), we must not forget to check whether $G(E)$ is {\it a sharply peaked function such as a Gaussian}.

Now, consider a particle tunneling through two equal successive rectangular barriers with height $V_0$ and width $a$:
\begin{equation}
V(x)\equiv V_0\left[ \theta (x)\theta (a-x)+\theta (x-L)\theta (L+a-x)\right] 
\end{equation}
where $0<a<L$, and $L-a$ is the distance between the barriers. About ten years ago, Olkhovsky, Recami and Salesi \cite{ORS} claimed that the tunneling time does not depend on the distance between the two barriers\footnote{ For an arbitrary number of barriers, see \cite{Esp}.}. However, we conclude that:
\begin{enumerate}
\item Although the SPM is used in the essential part of their reasoning in \cite{ORS}Aits applicability is questionable. The doubt arises in that their use of it is essentially the same as the incorrect use exemplified above. 
\item When we correctly apply the SPM to the wave packet which describes the particle, the tunneling time {\it does} in fact depend on the distance between the two barriers.
\item Contrary to our predictions, Longhi, Laporta, Belmonte and Recami \cite{LLBR} claim to have confirmed with experimental evidence the theoretical predictions obtained in Ref. \cite{ORS}. If this independence of the barrier separation is indeed confirmed, we are faced with a serious contradiction between our theory and physical reality. 
\end{enumerate}

First, consider the one-dimensional Schr{\" o}dinger equation
\begin{equation}
\left[ {1\over {2m}}{{d^2}\over {dx^2}}+(E-V(x))\right] \psi _E(x)=0.
\end{equation}
If we consider the plane wave $e^{ikx}$ with $k={\sqrt {2mE}}$ as the incident wave to the left of the potential, then we have 
\begin{equation}
\psi _E(x)=T(E)e^{ikx} \hskip 1cm (x>L+a)
\end{equation}
as the transmitted wave to the right of the potential, where the coefficient $T(E)$ has the form (see p.52 of Ref. \cite{Win}) 
\begin{equation}
T(E)={{e^{-ik(L+a)}}\over {(\cosh \chi a+i\Delta _-\sinh \chi a)^2e^{-ik(L-a)}+(\Delta _+\sinh \chi a)^2e^{ik(L-a)}}} \label{eq:Tnokatachi}
\end{equation}
with 
$$\chi ={\sqrt{2m(V_0-E)}}$$
and 
$$\Delta _+={1\over 2}\left( {\chi \over k}+{k\over \chi }\right) , \hskip 5mm \Delta _-={1\over 2}\left( {\chi \over k}-{k\over \chi }\right) .$$

Integrating over a band of stationary states with different energies, we construct a spatially localized incident wave packet
\begin{equation}
\int g(E)e^{ikx-iEt}dE,
\end{equation}
for which the transmitted wave has the form
\begin{equation}
\int g(E)T(E)e^{ikx-iEt}dE. \label{eq:transmittedpacket}
\end{equation}
In the following, $g(E)$ is assumed to be a Gaussian distribution given by
\begin{equation}
g(E)={1\over {{\sqrt{2\pi }}\delta }}e^{-(E-E_0)^2/2\delta ^2}, \label{eq:GEnosiki}
\end{equation}
where $E_0$ is within the interval $(0, V_0)$ and $\delta $ is rather small, such that the majority of the energy distribution is contained in $(0, V_0)$.

The authors of \cite{ORS} restrict themselves to the case where $a$ is large enough (and $\chi $ not too small) that one can assume $\chi a\rightarrow \infty $, and they obtain 
\begin{equation}
T(E)\rightarrow e^{-2\chi a}A{{-4ik\chi }\over {(ik-\chi )^2}}e^{-ik(L+a)} \label{eq:Tnozenkinke}
\end{equation}
where
\begin{equation}
A={{2k\chi }\over {2k\chi \cos k(L-a)+(\chi ^2-k^2)\sin k(L-a)}}.
\end{equation}
After confirming only that $A$ is real, they immediately conclude that we can derive that the {\it tunneling time} 
\begin{eqnarray}
\tau ^{\rm ph}_{\rm tun}&=&{{\partial \arg\left[ T(E)e^{ik(L+a)}\right] }\over {\partial E}}={\partial \over {\partial E}}\arg\left[ {{-4ik\chi }\over {(ik-\chi )^2}}\right] \nonumber \\
&=&{{\partial }\over {\partial E}}\arctan \left[ {{k^2-\chi ^2}\over {2k\chi }}\right] ={{2m}\over {k\chi }}, \label{eq:Recamitunn}
\end{eqnarray}
while depending on the energy of the particle, does not depend on $L+a$ (being actually independent of both $a$ and $L$). (See p.882 of Ref. \cite{ORS}.) 

If the term {\it tunneling time} means the time when the peak of the transmitted wave emerges at the point $x=L+a$, then their conclusion seems to be in doubt. The reason is as follows: In the limit $\chi a\rightarrow \infty $, the transmitted wave given by Eq. (\ref{eq:transmittedpacket}) has the form 
\begin{eqnarray}
&&\int g(E)T(E)e^{ikx-iEt}dE=\int g(E)e^{-2\chi a}A{{-4ik\chi }\over {(ik-\chi )^2}}e^{i\{ k(x-L-a)-Et\} }dE \nonumber \\
&&\hskip 2cm =\int g(E)e^{-2\chi a}A{{4k\chi }\over {k^2+\chi ^2}}e^{i\{ k(x-L-a)+\theta -Et\} }dE, 
\end{eqnarray}
where 
\begin{equation}
{{-4ik\chi }\over {(ik-\chi )^2}}={{4k\chi }\over {k^2+\chi ^2}}e^{i\theta }, \hskip 5mm \theta =\arctan {{k^2-\chi ^2}\over {2k\chi }}.
\end{equation}

{\it If the function} 
\begin{equation}
g(E)e^{-2\chi a}A{{4k\chi }\over {k^2+\chi ^2}} \label{eq:RecaminoG}
\end{equation}
{\it is of Gaussian type}, then using the SPM we can conclude that the position of the peak of the transmitted wave is given by 
\begin{equation}
{{dk}\over {dE}}(x-L-a)+{{d\theta }\over {dE}}-t=0.
\end{equation}
This implies that the peak emerges at $x=L+a$ at time 
\begin{equation}
t={{d\theta }\over {dE}}, \label{eq:dthetade}
\end{equation}
that is to say, the equation (\ref{eq:Recamitunn}) is reasonable.

\begin{figure}
\begin{center}
\includegraphics[height=7cm,width=14cm,keepaspectratio,clip]{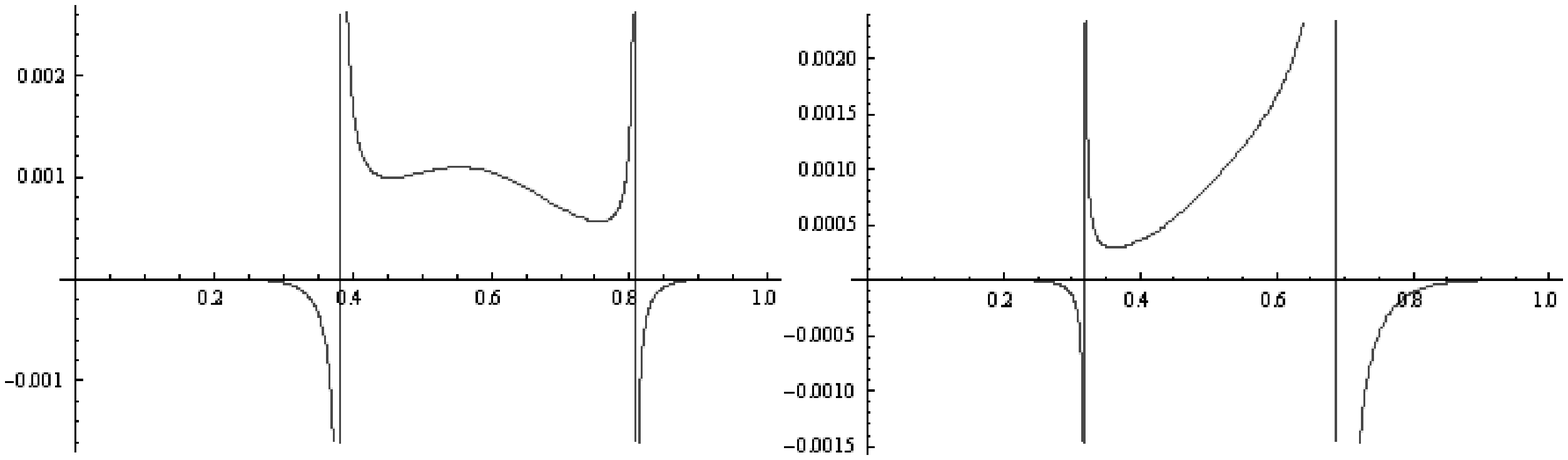}
\end{center}
\caption{ $g(E)e^{-2\chi a}A{{4k\chi }\over {k^2+\chi ^2}}$ given by Eq. (\ref{eq:RecaminoG})}
\label{figure:Recamifig}
\end{figure}

{\it But the function} (\ref{eq:RecaminoG}) {\it is far from Gaussian}: Fig. \ref{figure:Recamifig} shows the graph of (\ref{eq:RecaminoG}) with the width $a=5/{\sqrt{2mV_0}}$ and $E_0=V_0/2, \delta =V_0/10$. The scale on the horizontal axis indicates the ratio $E/V_0$. The left-hand figure is for the case $L-a=8/{\sqrt{2mV_0}}$ and the right-hand figure for $L-a=9/{\sqrt{2mV_0}}$. In that it has both positive and negative values and exhibits multiple extrema, the function (\ref{eq:RecaminoG}) is similar to the function (\ref{eq:myounagE}). Therefore, as exemplified at the beginning of this paper, there is a possibility that $\tau ^{\rm ph}_{\rm tun}$ given by Eq. (\ref{eq:Recamitunn}) does not represent the true tunneling time\footnote{ This kind of way of application of the SPM is seen frequently in literatures on this area.} . 

One further question arises in the reasoning given in \cite{ORS}: Even if the function (\ref{eq:RecaminoG}) is like a Gaussian centered on a certain point $E_1$, the tunneling time is not just the term $d\theta /dE$ of Eq. (\ref{eq:dthetade}). $d\theta /dE$ is a function of $E$, and the tunneling time $\tau ^{\rm ph}_{\rm tun}$ is obtained only after substituting $E_1$ for its variable $E$. Hence, even if the form of the function $d\theta /dE$ is independent of both $a$ and $L$ as shown by Eq. (\ref{eq:Recamitunn}), the tunneling time may depend on either $a$ or $L$ in the case that $E_1$ depends on $a$ or $L$. 

\begin{figure}
\begin{center}
\includegraphics[height=4cm,width=14cm,keepaspectratio,clip]{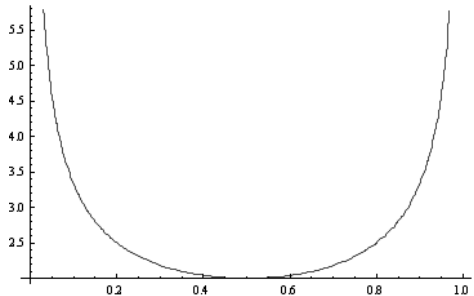}
\end{center}
\caption{$d\theta /dE=2m/k\chi $}
\label{figure:dvarphidE}
\end{figure}

Since the function (\ref{eq:RecaminoG}) is far from Gaussian, we cannot define $E_1$ as its mean energy. But, interpreting things broadly, we might try to define $E_1$ as the point at which the function tends to infinity. The value of $A$ is infinite if and only if its denominator 
$$2k\chi \cos k(L-a)+(\chi ^2-k^2)\sin k(L-a)$$
becomes null. Those null points vary as $L-a$ changes and the variation is rather sensitive to the change of $L-a$ as shown in Fig. \ref{figure:Recamifig}, and moreover those points occasionally come near the endpoints of the interval $(0, V_0)$. As we can see from Fig. \ref{figure:dvarphidE}\footnote{ The scales on the horizontal axis indicate the ratio $E/V_0$.}, the gradient of the function $d\theta /dE$ is very steep in the vicinity of those endpoints. Hence, there is a rather high possibility that the value of 
\begin{equation}
{{d\theta }\over {dE}}(E_1)
\end{equation}
depends on $L-a$.

Next, we show what we can obtain from a cautious application of the SPM to the transmitted wave (\ref{eq:transmittedpacket}): We do not use the approximation (\ref{eq:Tnozenkinke}), but use the original form (\ref{eq:Tnokatachi}). Introducing $r$ and $\phi $ by
$$\cosh \chi a+i\Delta _-\sinh \chi a=re^{-i\phi },$$
\begin{equation}
r={\sqrt{ 1+\Delta _+^2\sinh ^2\chi a}}, \hskip 5mm \phi =-\arctan (\Delta _-\tanh \chi a), \label{eq:rphinoteigi}
\end{equation}
we have 
\begin{eqnarray} 
&&T(E)={{e^{-ik(L+a)}}\over {(\cosh \chi a+i\Delta _-\sinh \chi a)^2e^{-ik(L-a)}+(\Delta _+\sinh \chi a)^2e^{ik(L-a)}}} \nonumber \\
&&=e^{-ik(L+a)}\left[ (re^{-i\phi })^2e^{-ik(L-a)}\left( 1+{{\left( \Delta _+\sinh \chi a\right) ^2}\over {(re^{-i\phi })^2}}e^{2ik(L-a)}\right) \right] ^{-1} \nonumber \\
&&=e^{-ik(L+a)}\sum _{n=0}^{\infty }(-1)^n{1\over {r^2}}\left( {{\Delta _+\sinh \chi a}\over r}\right) ^{2n}e^{i\{ (2n+1)k(L-a)+2(n+1)\phi \} }.
\end{eqnarray}

The transmitted wave (\ref{eq:transmittedpacket}) therefore can be written as 
\begin{eqnarray}
&&\sum _{n=0}^{\infty }(-1)^n\int {{g(E)}\over {r^2}}\left( {{\Delta _+\sinh \chi a}\over r}\right) ^{2n} \nonumber \\
&&\hskip 2cm \times e^{i\{ k(x-L-a)+(2n+1)k(L-a)+2(n+1)\phi -Et\} }dE, \label{eq:transmi}
\end{eqnarray}
that is to say, it is an infinite sum of the wave packets parameterized by $n=0, 1, 2, \cdots $ each of which has  
\begin{equation}
{{g(E)}\over {r^2}}\left( {{\Delta _+\sinh \chi a}\over r}\right) ^{2n} \hskip 5mm (n=0, 1, 2, \cdots ) \label{eq:namplittude}
\end{equation}
as its amplitude.  

\begin{figure}
\begin{center}
\includegraphics[height=5cm,width=14cm,keepaspectratio,clip]{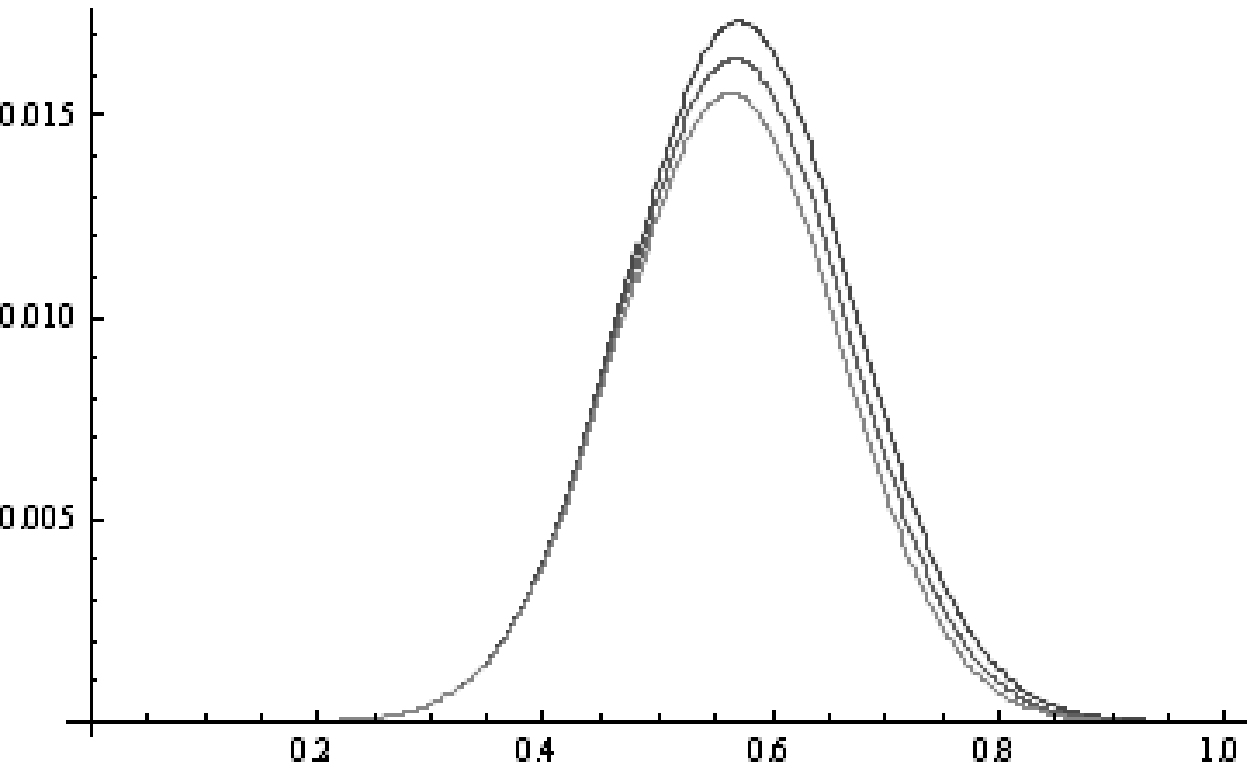}
\end{center}
\caption{ Graphs of Eq. (\ref{eq:namplittude})}
\label{figure:nthamplitude}
\end{figure}

Fig. \ref{figure:nthamplitude} shows the graphs of (\ref{eq:namplittude}) for the cases $n=0$ (upper), $10$ (middle) and $20$ (lower), where we set $E_0=V_0/2, \delta =V_0/10, a=5/{\sqrt {2mV_0}}$, and the scales on the horizontal axis represent the ratio $E/V_0$ \footnote{ Denote the transmission and the reflection coefficients for one potential barrier of width $a$ as $T_0$ and $R_0$, respectively. Then we have \cite{Win} 
$$T_0=(\cosh \chi a+i\Delta _-\sinh \chi a)^{-1}, \hskip 2mm R_0=-i\Delta _+\sinh \chi a(\cosh \chi a+i\Delta _-\sinh \chi a)^{-1}$$
and 
$$(\ref{eq:namplittude})=g(E)\vert T_0\vert ^2\vert R_0\vert ^{2n}.$$}. 

As shown in this figure, as far as $a$ does not exceed some critical value\footnote{ The transmission for a very thick barrier is over the top of the barrier for the most part, essentially not tunneling at all. This fact is stated as one of the conclusions of Hartman \cite{Hart}, and the critical value of $a$ depends on $E_0$ and $\delta $ determining the distribution $g(E)$ defined by Eq. (\ref{eq:GEnosiki}). We should pay attention to this fact when we treat the limit $a\rightarrow \infty $. (See \cite{KM}.)}, each of the amplitudes (\ref{eq:namplittude}) is {\it a sharply peaked function} centered at a certain point $E_1$ slightly higher than $E_0$. Therefore, the SPM can be applied to each term of the sum (\ref{eq:transmi}), and the position of the peak corresponding to $n$ is given by  
$${{dk}\over {dE}}(E_1)(x-L-a)+(2n+1)(L-a){{dk}\over {dE}}(E_1)+2(n+1){{d\phi }\over {dE}}(E_1)-t=0.$$
Hence, the $n$-th peak of the transmitted wave emerges at $x=L+a$ at
\begin{equation}
t=(2n+1)(L-a){{dk}\over {dE}}(E_1)+2(n+1){{d\phi }\over {dE}}(E_1). \label{eq:tnoatai}
\end{equation}

Since $L$ is not included in (\ref{eq:namplittude}), $E_1$ does not depend on $L$, and $dk/dE(E_1)$ and $d\phi /dE(E_1)$ are hence independent of $L$. Thus the time $t$ given by Eq. (\ref{eq:tnoatai}) is in proportion to $L-a$ for each $a$ and $n$.

Define $v$ and $\tau $ by
\begin{equation}
{1\over v}={{dk}\over {dE}}(E_1)={m\over {\sqrt{2mE_1}}}, \hskip 5mm \tau ={{d\phi }\over {dE}}(E_1),
\end{equation}
where (see p.9 of Ref. \cite{Win})
\begin{eqnarray}
&&{{d\phi }\over {dE}}={{ma}\over {2k}}\left( 1+\Delta _-^2\tanh ^2\chi a\right) ^{-1}  \nonumber \\
&&\hskip 1cm \times \Bigl[ \left( {k\over \chi }+{\chi \over k}\right) ^2{{\tanh \chi a}\over {\chi a}}-\left( {{k^2}\over {\chi ^2}}-1\right) {1\over {\cosh ^2\chi a}}\Bigr] 
\end{eqnarray}
and 
\begin{equation}
\tau ={{d\phi }\over {dE}}(E_1)\approx {{2m}\over {k(E_1)\chi (E_1)}}.
\end{equation}
The time $\tau $ is the {\it tunneling time} of a particle tunneling through one barrier, and it is independent of the barrier width under the condition, as shown by Hartman \cite{Hart}\footnote{ In his terminology $\tau =\delta t_3$.}, that the barrier is neither very thin nor very thick. 

Eq. (\ref{eq:tnoatai}) means that the peak of the $n$-th component of the transmitted wave (\ref{eq:transmi}) emerges at $x=L+a$ at
\begin{equation}
t=(2n+1){{L-a}\over v}+2(n+1)\tau . \label{eq:nnpeak}
\end{equation}
These times {\it do} thus depend on the distance $L-a$.

If we consider the behavior of the wave 
\begin{equation}
\int g(E)\psi _E(x)e^{-iEt}dE \label{eq:geeite}
\end{equation}
in the regions $x<0$ and $a<x<L$, we can gain a better understanding of why the transmitted wave has multiple peaks: We define the coefficients $R(E), \alpha (E)$ and $\beta (E)$ by
\begin{equation}
\psi _E(x)=\cases{ e^{ikx}+R(E)e^{-ikx} & $x<0$, \cr 
\alpha (E)e^{ik(x-a)}+\beta (E)e^{-ik(x-a)} & $a<x<L$. \cr } 
\end{equation}

Simple calculations show that 
\begin{eqnarray*}
&&R(E)={{-i\Delta _+\sinh \chi a}
\over {\left( \cosh \chi a+i\Delta _-\sinh \chi a\right) ^2e^{-ik(L-a)}+\Delta _+^2\sinh ^2\chi ae^{ik(L-a)}}}\\
&&\times \left[ \left( \cosh \chi a-i\Delta _-\sinh \chi a\right) e^{ik(L-a)}+\left( \cosh \chi a+i\Delta _-\sinh \chi a\right) e^{-ik(L-a)}\right] .
\end{eqnarray*}
Using $r, \phi $ introduced by Eq. (\ref{eq:rphinoteigi}), we have 
\begin{eqnarray}
&&R(E)=-i\Delta _+\sinh \chi a\left[ (re^{-i\phi })^2e^{-ik(L-a)}\left( 1+{{\Delta _+^2\sinh ^2\chi a}\over {(re^{-i\phi })^2}}e^{2ik(L-a)}\right) \right] ^{-1}\nonumber \\
&&\hskip 3cm \times \left[ re^{ik(L-a)+i\phi }+re^{-ik(L-a)-i\phi }\right] \nonumber \\
&&=-ie^{i\phi }\sum _{n=0}^{\infty }(-1)^n\left( {{\Delta _+\sinh \chi a}\over r}\right) ^{2n+1} \nonumber \\
&&\hskip 2cm \times \left[ e^{i\{ 2(n+1)k(L-a)+2(n+1)\phi \} }+e^{i\{ 2nk(L-a)+2n\phi \} }\right] \nonumber \\
&&=-i{{\Delta _+\sinh \chi a}\over r}e^{i\phi }\nonumber \\
&&\hskip 1cm +i\sum _{n=1}^{\infty }(-1)^n{1\over {r^2}}\left( {{\Delta _+\sinh \chi a}\over r}\right) ^{2n-1}e^{i\{ 2nk(L-a)+(2n+1)\phi \} }. \label{eq:REnokatachi} 
\end{eqnarray}

The decomposition (\ref{eq:REnokatachi}) implies that the reflected part of the wave (\ref{eq:geeite}) is an infinite sum 
\begin{eqnarray*}
&&\int g(E)R(E)e^{i\{ -kx-Et\} }dE=-i\int g(E){{\Delta _+\sinh \chi a}\over r}e^{i\{ -kx+\phi -Et\} }dE \nonumber \\
&&+i\sum _{n=1}^{\infty }(-1)^n\int {{g(E)}\over {r^2}}\left( {{\Delta _+\sinh \chi a}\over r}\right) ^{2n-1}e^{i\{ -kx+2nk(L-a)+(2n+1)\phi -Et\} }dE, 
\end{eqnarray*}
where the amplitudes 
\begin{equation}
g(E){{\Delta _+\sinh \chi a}\over r}\hskip 3mm  {\rm and}\hskip 3mm  {{g(E)}\over {r^2}}\left( {{\Delta _+\sinh \chi a}\over r}\right) ^{2n-1}\hskip 2mm (n=1, 2, \cdots )
\end{equation}
are {\it all Gaussian type} as shown in Fig. \ref{figure:nthamplitude}. The SPM, therefore, can be applied and shows us the position of the peak of each term of the sum above; the first peak is given by 
\begin{equation}
-x/v+\tau -t=0
\end{equation}
and the subsequent peaks are given by
\begin{equation}
-x/v+2n(L-a)/v+(2n+1)\tau -t=0\hskip 1cm (n=1, 2, \cdots ).
\end{equation}
In particular, the departure times from $x=0$ of those peaks are given by 
\begin{equation}
t=2n(L-a)/v+(2n+1)\tau \hskip 1cm (n=0, 1, 2, \cdots ).
\end{equation}
Consecutive departure times are separated by $2(L-a)/v+2\tau $.

A similar discussion for the behavior of the wave (\ref{eq:geeite}) in the region $a<x<L$ is possible: The coefficients $\alpha (E), \beta (E)$ are 
\begin{eqnarray*}
\alpha (E)&=&{{(\cosh \chi a+i\Delta _-\sinh \chi a)e^{-ik(L-a)}}
\over {\left( \cosh \chi a+i\Delta _-\sinh \chi a\right) ^2e^{-ik(L-a)}+\Delta _+^2\sinh ^2\chi ae^{ik(L-a)}}} \\
&=&\sum _{n=0}^{\infty }(-1)^n{1\over r}\left( {{\Delta _+\sinh \chi a}\over r}\right) ^{2n}e^{i\{ 2nk(L-a)+(2n+1)\phi \} },  \\
\beta (E)&=&{{-i\Delta _+\sinh \chi ae^{ik(L-a)}}
\over {\left( \cosh \chi a+i\Delta _-\sinh \chi a\right) ^2e^{-ik(L-a)}+\Delta _+^2\sinh ^2\chi ae^{ik(L-a)}}} \\
&=&-i\sum _{n=0}^{\infty }(-1)^n{1\over r}\left( {{\Delta _+\sinh \chi a}\over r}\right) ^{2n+1}e^{i\{ 2(n+1)k(L-a)+2(n+1)\phi \} },  
\end{eqnarray*}
which indicate that the multiple peaks of right$-$ and left$-$ going waves emerge in the region $a<x<L$. The right$-$ going peaks depart from $x=a$ at 
\begin{equation}
t=2n(L-a)/v+(2n+1)\tau \hskip 1cm (n=0, 1, 2, \cdots )
\end{equation}
and the left$-$ going peaks arrive at $x=a$ at
\begin{equation}
t=2(n+1)(L-a)/v+2(n+1)\tau \hskip 1cm (n=0, 1, 2, \cdots ).
\end{equation}

\begin{figure}
\begin{center}
\includegraphics[height=7cm,width=14cm,keepaspectratio,clip]{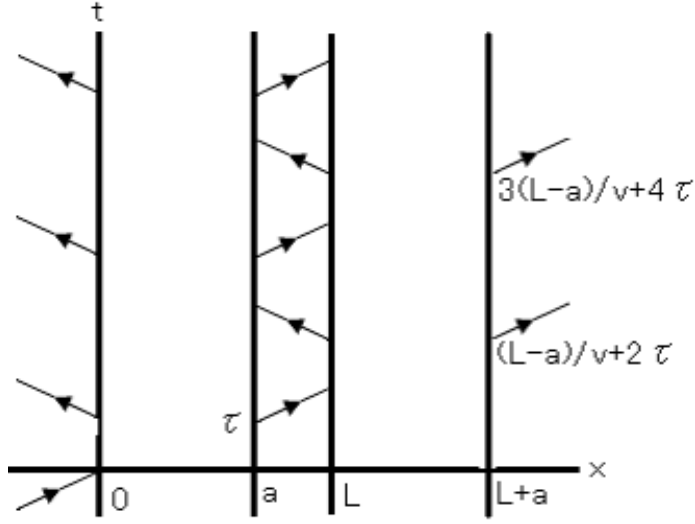}
\end{center}
\caption{Motion of the peaks}
\label{figure:Wpfig1}
\end{figure}

The motions of the peaks thus obtained can be represented as in Fig. \ref{figure:Wpfig1}, from which we can understand that the multiple reflections in the region $a<x<L$ cause the existence of the multiple peaks in the transmitted wave in the region $x>L+a$.

De Leo and Rotelli \cite{LeoRo} state that the generalized Hartman effect represents an example of an ambiguity in the use of the SPM. They state that although the SPM concludes the existence of a generalized Hartman effect if the tranmitted wave is assumed to have a single peak, the same method concludes an alternative result if the transmitted wave is assumed to have multiple peaks. Admitting the posibility of multiple outgoing peaks, they demonstrate the same formula as Eq. (\ref{eq:nnpeak}). 

Following the results obtained in this paper, we have to emphasize two points: (a) The generalized Hartman effect is {\it not} an example of an ambiguity in the use of the SPM, but rather an example of a mathematically incorrect application. There is no ambiguity in the use of the SPM. (b) The fact that the transmitted wave has multiple peaks is not an assumption. It is an inevitable result of a mathematically correct application of the SPM.  

Finally, we have to turn our attention to a report written by Longhi, Laporta, Belmonte and Recami \cite{LLBR} in which tunneling optical pulses at $1.5$ $\mu $m wavelength through double-barrier periodic fiber Bragg gratings is experimentally investigated. They claim that the transit time is paradoxically short and almost independent of the barrier distance. They interpret their result as providing, in the optical context, experimental evidence of the analogous phenomenon in quantum mechanics of nonresonant superluminal tunneling of particles across two successive potential barriers. 

As we have discussed above, the cautious application of the SPM shows that the transmitted wave has an infinite number of peaks and that the time each peak emerges at $x=L+a$ depends on the distance $L-a$. Therefore, if the interpretation of the authors of \cite{LLBR} is accepted, we are faced with a serious contradiction between theory and physical reality. Resolution of this contradiction is very important in order to encourage steady progress in this research area.

\vfill

\begin{thebibliography}{99}
\bibitem{ORS} V. S. Olkhovsky, E. Recami, and G. Salesi, Europhys. Lett. {\bf 57} (2002) 879.
\bibitem{Esp} S. Esposito, Phys. Rev. E {\bf 67} (2003) 016609; Y. Aharonov, N. Erez, and B. Reznik, Phys. Rev. A {\bf 65} (2002) 052124.
\bibitem{LLBR} S. Longhi, P. Laporta, M. Belmonte, and E. Recami, Phys. Rev. E {\bf 65} (2002) 046610. 
\bibitem{Win} H. G. Winful, Phys. Rep. {\bf 436} (2006) 1.
\bibitem{Hart} T. E. Hartman, J. Appl. Phys. {\bf 33} (1962) 3427.
\bibitem{KM} S. Kudaka, and S. Matsumoto, in preparation; A. E. Bernardini, Ann. Phys. {\bf 324} (2009) 1303; G. Privitera, G. Salesi, V. S. Olkhovsky and E. Recami, Riv. Nuovo Cimento {\bf 26} (2003) 1.
\bibitem{LeoRo} S. De Leo, P. R. Rotelli, Phys. Lett. A {\bf 342} (2005) 294.
\end{thebibliography}
\end{document}